\documentclass[12pt,preprint]{emulateapj}
\usepackage{graphics}
\usepackage{subfigure}
\usepackage{hyperref}
\usepackage{graphicx}
\usepackage{gensymb}

\usepackage{float}

\usepackage{amsmath}
\usepackage{natbib}
\usepackage{float}
\usepackage{adjustbox}
\usepackage{booktabs}
\usepackage{amsmath}

\begin{document}

\title{From Exoplanets to Quasars: Detection of Potential Damped Lyman Alpha Absorbing Galaxies using Angular Differential Imaging}

\author{Mara Johnson-Groh\altaffilmark{1}, Christian Marois\altaffilmark{2,1}, and Sara L. Ellison\altaffilmark{1}}

\affil{$^{1}$Department of Astronomy, University of Victoria, 3800 Finnerty Rd, BC V8P 5C2, Canada; mara@uvic.ca }
\affil{$^{2}$National Research Council Canada, Herzberg, 5071 West Saanich Rd, Victoria, BC, V9E 2E7, Canada}


\begin{abstract}

The advantages of angular differential imaging (ADI) has been previously untested in imaging the host galaxies of damped Lyman $\alpha$ (DLA) systems.   In this pilot study, we present the first application of ADI to directly imaging the host galaxy of the DLA seen towards the quasar J1431+3952. K-band imaging of the field surrounding J1431+3952 was obtained on the Gemini North telescope with the adaptive optics system and a laser guide star.  We computed a sensitivity curve that demonstrates the sensitivity of our observations as a function of K-band magnitude, impact parameter and DLA angular size.  For an impact parameter of 0.5" (3.4 kpc at the redshift of the absorber) our mass sensitivity is log (M$_{\star}$/M$_{\odot}) \sim$ 9.2 and drops to $\sim$ 9.0 at separations beyond $\sim$ 6 kpc for the smallest size model galaxy.  Three candidate galaxies are identified within 5". Stellar masses were computed from the K-band photometry yielding values of log (M$_{\star}$/M$_{\odot}) \sim$ 9.9, 9.7 and 11.1 respectively.  The likely identification of the absorbing galaxy is discussed, and we conclude that the galaxy with the largest impact parameter and highest stellar mass is unlikely to be the host, based on its inconsistency with the N(HI) impact parameter relation and inconsistent photometric redshift.   Whilst we cannot distinguish between the remaining two candidates as the DLA host, we note that despite the low spin temperature and relatively high metallicity of the DLA, the host does not appear to be a particularly luminous (high mass) galaxy.

\end{abstract}

\section{INTRODUCTION}\label{sec:intro}  

When a quasar's line of sight is intercepted by intervening gas, the resulting absorption lines provide information about the chemical and physical properties in the early universe.  The highest column density absorbers with log N(HI) $\ge$ 20.3 are the so-called damped Lyman $\alpha$ systems (DLAs), which have proven to be powerful probes of high redshift galaxies \citep{wolfe2}.   Given the slope of the column density distribution function of high redshift absorption line systems \citep{tytler}, it is the DLAs, rather than the more numerous Ly$\alpha$ forest clouds, or Lyman limit systems, that contain the bulk of the neutral gas available for star formation.  The abundance of the DLA atomic gas reservoir is little changed over many Gyrs at intermediate and high redshifts \citep[e.g.][]{chrighton,sanchez-ramirez,neeleman} indicating that DLAs offer a fertile location for star formation over a significant fraction of cosmic time.  Combined with the chemical enrichment associated with DLAs, which manifest a wide range of metallicities from a less than 1/100 to in excess Z$_{\odot}$ metallicity \citep[e.g.][and references therein]{berg}, we expect these high column density absorbers to represent a broad cross section of galaxies and hence provide a window into galaxy evolution at these epochs.  Indeed, there is a large body of research spanning the last two decades and beyond that have studied many aspects of DLAs, ranging from their elemental ratios \citep[e.g.][]{pettini00,prochaska,dessauges-zavadsky}, molecular fraction \citep[e.g.][]{ledoux03,noterdaeme}, kinematics \citep[e.g.][]{prochaskawolfe,ledoux06,neeleman13}, ionization properties \citep[e.g.][]{vladilo01,milutinovic} and dust depletion in the interstellar medium \citep[ISM, e.g.][]{pettini94,pettini97,vladilo11,murphy}.  Despite these advances, some of the most fundamental properties of the absorbing galaxies, such as their luminosities, stellar masses and morphologies remain unknown for the vast majority of the DLA population.   For many years, most DLAs with identified host galaxies were at relatively low-to-intermediate redshift \citep[e.g.][]{rao, bowen,chen,lebrun},  but there are now a growing number detected at $z>2$ \citep[e.g.][]{djorgovski,weatherley,fynbo,krogager,peroux12,krogager13,kashikawa,jorgenson14,hartoog,mawatari,peroux16}.  Nonetheless, the identification of wholesale numbers of host galaxies for DLAs remains one of the outstanding challenges in the field.

The fundamental challenge for the identification of a DLA host galaxy is the overwhelming brightness of the quasar relative to the galaxy, which is expected to be found at low impact parameter from the QSO \citep{rao03}.  A few methods have been developed to circumvent the blinding quasar light, and to allow the observer to search for galaxy hosts close to the QSO line of sight.  One such method is the 'Double-DLA' technique \citep{steidel92,fum,omeara06}, in which a QSO exhibits multiple high column density absorbers.  The higher redshift system is used as a natural blocking filter to eliminate (rest-frame) far ultraviolet emission from the quasar so that the continuum emission from the lower redshift DLA can be directly measured.  An alternative approach has been to search for absorbers in the spectra of the optical afterglow of gamma-ray bursts (GRBs) \citep[e.g.][]{vreeswijk,chen05,prochaska07}.  GRB afterglows can shine brighter than quasars, but they fade rapidly \citep{kann}.  Spectra taken along the sightline of the GRB can reveal intervining DLA, sub-DLA and other absorbing systems \citep{schulze}.  After the afterglow has faded, follow up imaging and spectroscopy can be used to identify the host galaxy \citep[e.g.][]{masetti, pollack,vree,schulze,ellison06,chen09,chen10}.  Again, this technique is limited to a relatively small number of absorbers.  Finally, spectroscopy with integral field units (IFUs) has been used to successfully identify DLA hosts at z $\sim$ $1-2$ \citep[e.g.][]{peroux,bouche}.  This technique uses narrow band images generated from IFU data cubes to search for H$\alpha$ or other emission lines along quasar sightlines, and is applicable when the DLA redshift optimally places emission lines between the bright sky lines in the IR.

In the pursuit of identifying DLA hosts, adaptive optics (AO) seems a natural facilitator.  AO has an obvious application in the search for DLA host galaxies, thanks to the improved diffraction-limited angular resolution and deep contrast close to the quasar that can be achieved.  Despite its obvious benefits, relatively few AO-aided searches for DLA hosts outside IFU useage have been attempted in the past \citep{chun06,chun10}.   Part of the historical challenge of AO observations has been the limited sky coverage of natural guide stars.  Moreover,  AO imaging techniques require fairly complex analysis methods, to deal with a highly time and space sensitive point spread function (PSF).  Fortunately, significant progress has been made developing data reduction techniques, and the introduction of laser guide star technology opens the night sky to the application of AO.

In this paper we consider the application of a particular AO technique called angular differential imaging (ADI), which was developed for directly imaging exoplanets \citep{marois06}.  The main technical feature of the ADI technique is the disabling of the instrument rotator during observations, so that the field-of-view rotates around a central axis during the set of exposures.  Given enough field-of-view rotation, these images can be combined to create a reference PSF and suppress quasi-static speckles by up to two orders of magnitude.  The reference PSF removes off-axis light which increases the sensitivity, allowing for the detection of fainter objects at lower impact parameters.  Previous AO imaging of DLAs have largely used multi-step methods of azimuthal PSF subtraction \citep{chun10,chun06}.  Applying ADI to AO imaging of DLAs can greatly simplify the reduction process and potentially improve the limits of detection.  

This paper is organized as follows: Section~\ref{methods} describes the DLA that is the target of this pilot study, including details of14 the methods of observation (\ref{obsv}), data reduction (\ref{redu}) and PSF subtraction.  In Section \ref{results} we describe the determination of candidate positions and magnitudes (\ref{simul}), our calculations of stellar mass (\ref{mass}), and the detection limits of this study (\ref{curves}). Section~\ref{disc} discusses candidate DLA host galaxies in the context of known scaling relations and directions for future work. Section~\ref{conc} summarizes our conclusions.  Throughout this paper, a $\Lambda$ cold dark matter cosmology is assumed with $H_0$ = 68 km s$^{-1}$ Mpc$^{-1}$,  $\Omega_\Lambda$ = 0.70 and $\Omega_M$ = 0.30.


\section{METHODS}\label{methods}

\subsection{Target selection and overview}

AO requires a bright point source (usually a star) close to the target in order to compute corrections to the incoming wavefront.  When there is no star bright enough close to the target, it is necessary to use a laser guide star (LGS).  The LGS works by using a 589nm laser to excite the sodium layer at an altitude of $\sim$90km which the telescope can then use for high-order atmosphere corrections.  Although the LGS is useful in fields without bright guide stars, a source is still needed for tip/tilt corrections.  If the target is bright enough (R-band apparent magnitude m$_R$\textless17.5), it can be used for tip/tilt corrections.  For this pilot study, we therefore selected a relatively bright QSO with a known DLA: J1431+3952  (z$_{em}$=1.215, m$_K$=14.03, m$_R$=16.07).  The only previous attempts to identify the galaxy counterpart of this DLA relied on shallow, relatively poor seeing quality\footnote{The typical seeing for SDSS, given by the PSF width, is 1.43" \citep{sdss}.} Sloan Digital Sky Survey (SDSS) imaging.  Both Ellison et al. (2012) and Zwaan et al. (2015) identified a galaxy at an impact parameter of $\sim$ 5 arcsec; although the relative proximity makes this galaxy an appealing host galaxy candidate, its photometric redshift was determined by Ellison et al. (2012) to be $z=0.08\pm0.02$, inconsistent with the absorption redshift  ($z_{abs}=0.602$).  

In addition to the advantageous brightness of the background QSO, we selected J1431+3952 for our pilot observations because of extensive characterization of its DLA by UV, optical and radio spectroscopy.  Initially selected as a candidate DLA based on its strong MgII lines, \citet{ellison12} used UV spectroscopy with the Cosmic Origins Spectrograph (COS) on the Hubble Space Telescope (HST) to determine an HI column density of log N(HI) = 21.2$\pm 0.1$.  Thanks to its radio brightness, J1431+3952 has also been the target of 21cm absorption measurements (Ellison et al. 2012; Zwaan et al. 2015). By combining the 21cm optical depth and N(HI) determined from Ly$\alpha$, it is possible to determine the spin temperature (T$_s$), which indicates the fraction of warm and cool neutral medium of the intervening DLA.  The majority of DLAs for which spin temperature measurements exist exhibit values of T$_s$ $>$ 1000 K, with relatively few low spin temperature measurements, particularly at high redshifts   \citep[e.g.][for a rare example of a low T$_s$, high $z$ absorber]{kanekar14b,kanekar13,kanekar,york}. Ellison et al. (2012) determined that the DLA towards J1431+3952 exhibits one of the lowest spin temperatures yet measured (T$_s$ = 90$\pm$23 K), and Zwaan et al. (2015) report an even lower value  (T$_s$ = 65$\pm$17 K)\footnote{Likely due to
a larger measured integrated optical depth of the abosrption feature, although \citet{zwaan} do also estimate a larger covering factor 3 times larger than \citet{ellison12}.}.  An anti-correlation between the gas phase metallicity and T$_s$ has been proposed by \citet{kanekar}, whereby the abundance of metals allow effective cooling in the ISM \citep{kanekar01}.  The most recent compilations of abundance and spin temperature measurements seem to support this anti-correlation \citep{kanekar14}, although sample sizes remain small.  The general association between high elemental abundances and low T$_s$ is observed in the DLA towards J1431+3952:  Ellison et al. (2012) determine a relatively high (compared with the general DLA population, e.g. Berg et al. 2015) metallicity of [Zn/H] $ = -0.80$$\pm$$0.13$.  

A further prediction of \citet{kanekar} is that low spin temperature DLAs should be hosted by relatively luminous galaxies.  A stellar mass--metallicity relation is observed over a wide range in redshifts, such that more metal rich galaxies are hosted by more massive galaxies  \citep[e.g.][]{tremonti,erb,zahid14}, which should in turn be more luminous.  Indeed, the small number of low spin temperature DLAs which have galaxy counterparts tend to be relatively luminous spirals \citep{kanekar02,kanekar03}.  However, very few DLAs have the full complement of data that permit an investigation between metallicity, spin temperature and identified galaxy counterpart to more completely characterize the inter-relation of these properties.  The selection of a low T$_s$ DLA, with high metallicity (that is hence predicted to be hosted by a relatively bright galaxy), and also towards a bright background QSO, therefore makes an excellent target for our pilot test of the ADI technique applied to the search for DLA hosts.  A complete summary of the DLA's properties can be found in Table~\ref{dlaprops}.

\begin{table}[H]
\centering
\centering\caption{Properties of the DLA associated with QSO J1431+3952}
\begin{tabular}{ccc}
    \toprule
    \midrule
    \bfseries Property & 
    \bfseries Value & 
    \bfseries Reference \\
    \midrule
z$_{abs}$ &0.60190 &1 \\
logN(HI)&21.2$\pm$0.1&1\\
{[Fe/H]}&$-$1.50$\pm$0.11&1\\
{[Zn/H]}&$-$0.80$\pm$0.13& 1\\
{[Cr/H]}&$-$1.31$\pm$0.13&1\\
{[Mn/H]}&$-$1.61$\pm$0.12&1\\
{[Ti/H]}&$-$1.41$\pm$0.12&1\\
T$_s$(K)& 90$\pm$23, 65$\pm$17&1,2\\
Covering Fraction &0.32,  0.95&1,2\\
    \midrule
    \toprule
\multicolumn{3}{c}{References: 1 - \citet{ellison12}; 2 - \citet{zwaan}. } \\
\multicolumn{3}{c}{Abundances determined using solar values} \\
\multicolumn{3}{c}{ from \citet{asplund}.} \\
\end{tabular}   
\label{dlaprops}
\end{table}

\subsection{Observations}\label{obsv}

The quasar J1431+3952 was observed using the f/32 camera on the Near InfraRed Imager and Spectrometer \citep[NIRI,][]{hodapp} at Gemini North over three nights in 2013 during program GN-2013A-Q-17. NIRI was used with the Gemini AO system, ALTAIR \citep[ALTtitude conjugate Adaptive optics for the InfraRed,][]{herriot}.  For the AO system, a LGS was used in conjunction with the quasar, the latter of which was sufficiently bright for tip/tilt corrections.  The NIRI pixel scale with the f/32 field lens is 0.0214 arcsec/pixel.  The typical Strehl ratio is $\sim$10\%.

The images were all taken in the K-short band ($1.99-2.30$ microns) with 60 second exposures of one coadd at the time when the object was closest to transit, so as to maximize the field-of-view rotation.  Images were obtained for three nights, however,  one night showed a strong elongation in the light from the quasar, most likely due to the errors locking the AO system, and so could not be used in this study.  Of the images used, 19 were from April 27, 2013 and 123 were from April 26, 2013.  These 142 images were combined using a signal to noise ratio weighting scheme as described in Section ~\ref{redu}.  A full list of observing conditions can be found in Table~\ref{observations}.
\begin{deluxetable}{ccc}
\tabletypesize{\scriptsize}
\tablewidth{0pt}
\setlength{\tabcolsep}{3pt}
\tablecaption{Observational Details}
\tablehead{{Date}&{26 April 2013} &{27 April 2013}}
\startdata
Number of Images & 123 & 19\\
FOV Rotation & 86.0\degree & 15.3$\degree$ \\
Average airmass & 1.13 & 1.07 \\
Atmospheric Seeing &  0.41$\pm$0.06 & 0.30$\pm$0.04\\
\enddata
\label{observations}
 \tablecomments{Atmospheric Seeing is listed as the average with the error representing the standard deviation in seeing during the course of the night.}
\end{deluxetable}

\subsection{Reduction Methods}\label{redu}

The images from the two nights in April were each processed separately using a routine developed specifically for this project.  A master dark image was made from a median combination of all dark images taken that night, and a master flat was constructed from the median of all the flat images, after they had been dark subtracted and flux normalized to one.  Each quasar science image was dark subtracted and divided by the master flat field image. All science images from April 26 had vertical striping in the lower left quadrant due to an error in detector readout.  Corrections were made by taking the first ten rows in each quadrant, calculating the median value of each column and subtracting that column median value from each row over the entire quadrant.  Similarly, the images from both nights had distinct horizontal readout bias from the detector chips.  This was corrected by taking the first hundred columns in each row, calculating the median value, and subtracting that median value from each pixel in the whole row.   Additional corrections were made for instrumental distortion using a distortion map by R. Galicher (private communication).  After corrections, the image was unsharp masked with a large median box of 50 $\times$ 50 pixels (1.07 $\times$ 1.07 arcseconds) to remove the background flux and yet large enough to avoid removing flux for a resolved and diffuse galaxy.  Each image was registered to the common centre with sub-pixel accuracy by fitting a gaussian to the quasar's PSF with the IDL routine \texttt{gauss2dfit.pro}.  Registration was executed with a custom IDL routine using the IDL function 'rot' with a cubic convolution interpolation method and an interpolation parameter of -0.5.

While very advanced PSF subtraction techniques exist, such as LOCI \citep{lafreniereb}, SOSIE \citep{marois10}, KLIP  \citep{soummer}, TLOCI \citep{marois14} and LLSG \citep{gomez}, we decided to use a simple subtraction algorithm to minimize self-subtraction since the DLA host is likely to be resolved.  For our method, a reference PSF for the quasar was created by taking the median of all the science images for each night.  This PSF was subtracted from each individual image, significantly removing the quasar's signal, and allowing the detection of low impact parameter galaxies.  ADI has the advantage of offering high correlation between PSFs in consecutive exposures, enabling a precise reconstruction of the overall PSF.  Any off-axis point-source light, such as light from a DLA host galaxy, will be mostly median averaged out in the reference PSF and will not be significantly subtracted from the final image. Extended objects, however, will be subject to some self-subtraction.  As the quasar image is not moving on the detector during the entire observing sequence, PSF subtraction simultaneously doubles for sky subtraction, resulting in an increased signal to noise over the whole image, not just at the centre where the PSF is subtracted.  This allows for substantial time savings during observations since sky images are not required.  It additionally can remove detector artifacts, which vary on timescales of hours, since the science images (that are also used for image subtraction) are taken when the detector artifacts are still highly coherent, not hours before or after the data is taken.  After PSF subtraction, the images were rotated to north up using the IDL function 'rot' which used the cubic convolution interpolation method with an interpolation parameter of -0.5. The rotated images were then median combined.  Since the two nights have different seeing and number of images, it is important to combine them in a way that weights the noise from each night to maximize the detection limit.  To achieve this, a final combined image was created from a weighted mean of the images from the two nights.  This was done by scaling the flux of the images to be the same and then combining the images with a weight determined by their signal to noise ratios:

\begin{equation}
image = \frac{(a*snr_a^2)+(b*snr_b^2)}{snr_a^2+snr_b^2},
\label{eq:weight}
\end{equation}

where $a$ refers to images from one night, and $b$ the other.  The signal to noise ratios for each image ($snr_a$ and $snr_b$) are calculated by dividing the maximum value of the quasar by the standard deviation in an empty annulus around the quasar. The SNR ratio for the 27th to the 26th was 0.599.

\section{Results}\label{results}

\begin{figure}[h]
  \centering
    \begin{tabular}[b]{@{}p{0.32\textwidth}@{}}
   \includegraphics[width=1.2\linewidth]{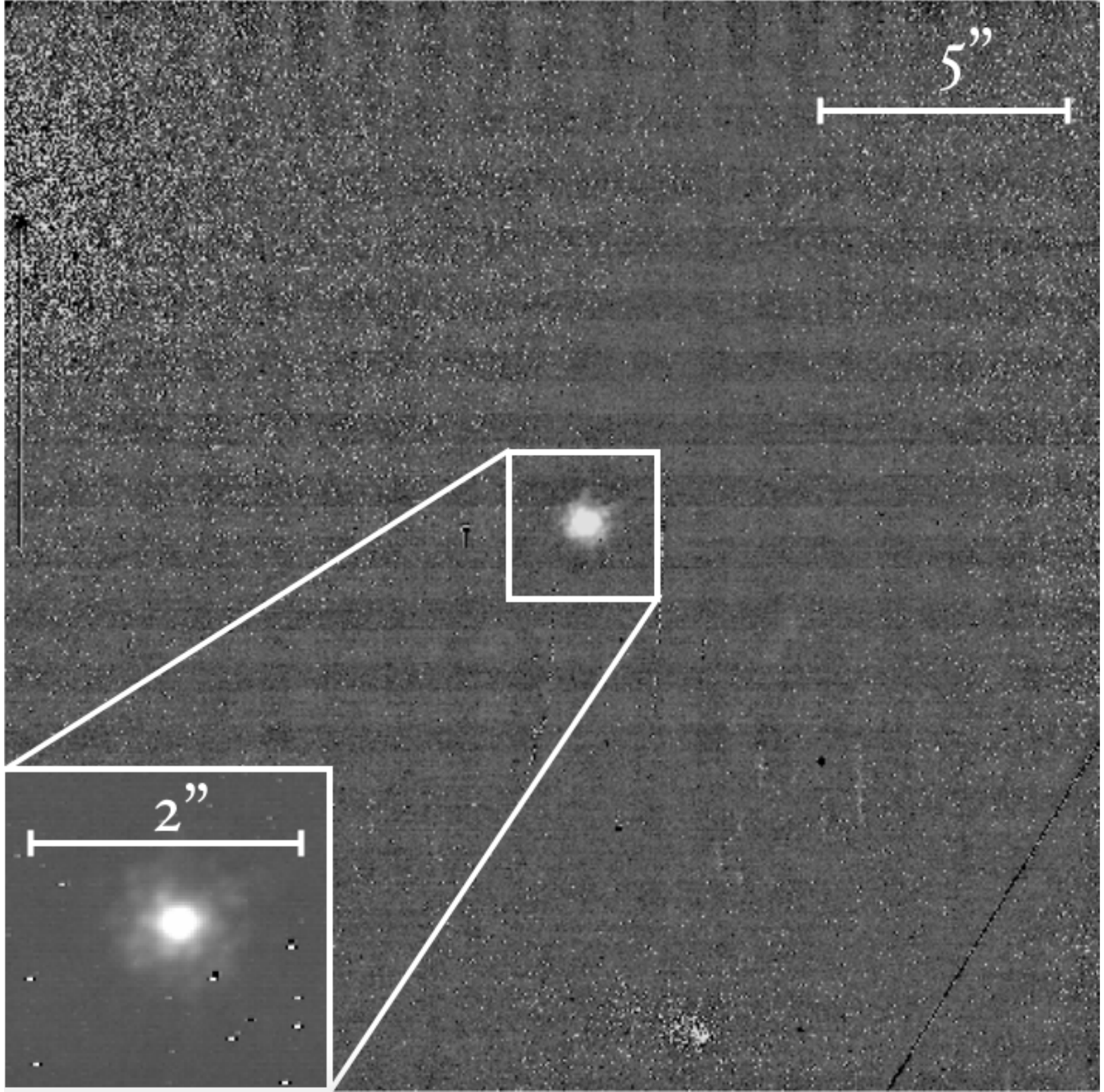} \\
    \centering\small (a) Quasar PSF  
  \end{tabular}%
  \begin{tabular}[b]{@{}p{0.36\textwidth}@{}}
   \includegraphics[width=1.2\linewidth]{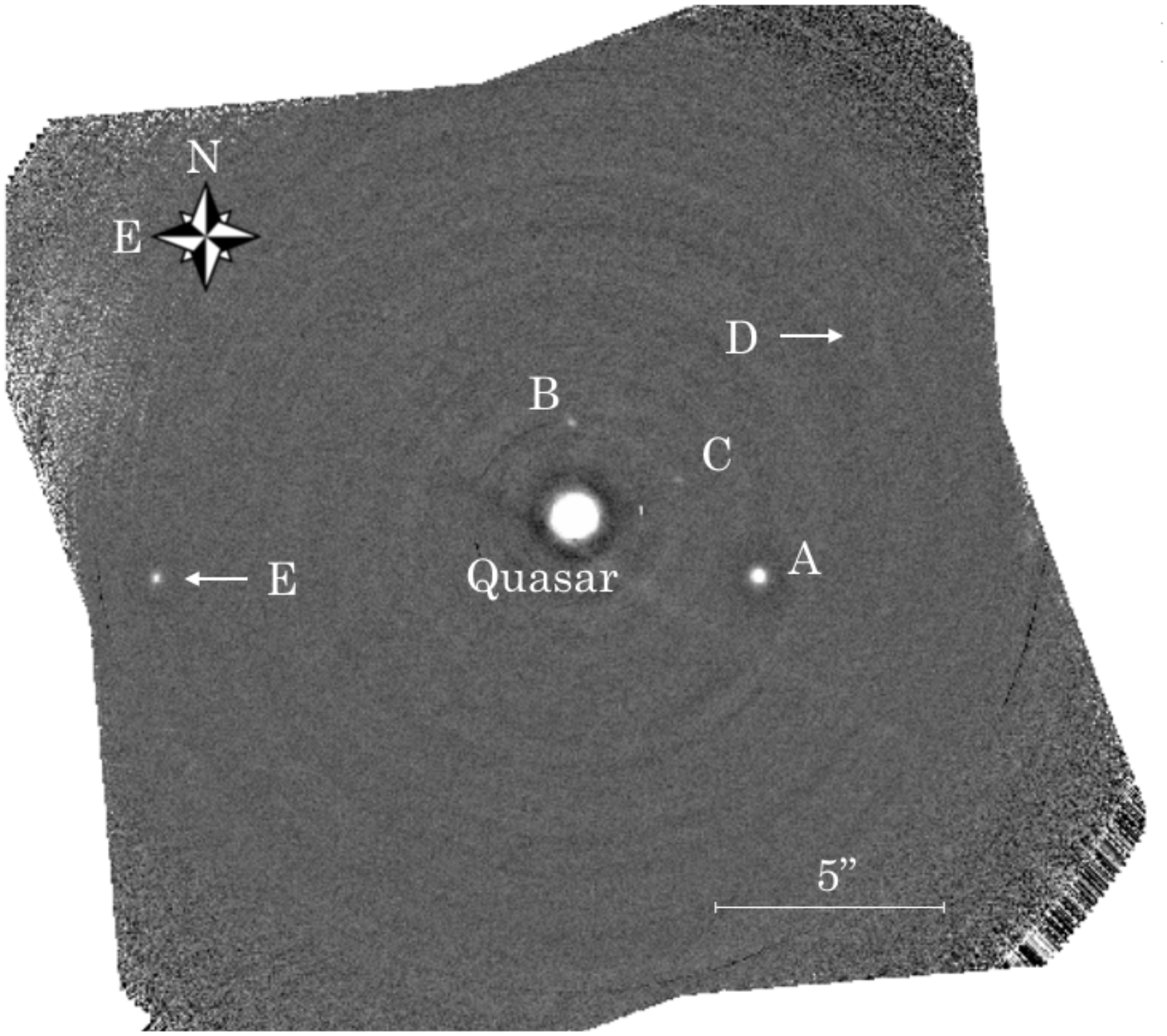} \\
    \centering\small (b)   Image without PSF subtraction
  \end{tabular}%
  \begin{tabular}[b]{@{}p{0.36\textwidth}@{}}
    \includegraphics[width=1.2\linewidth]{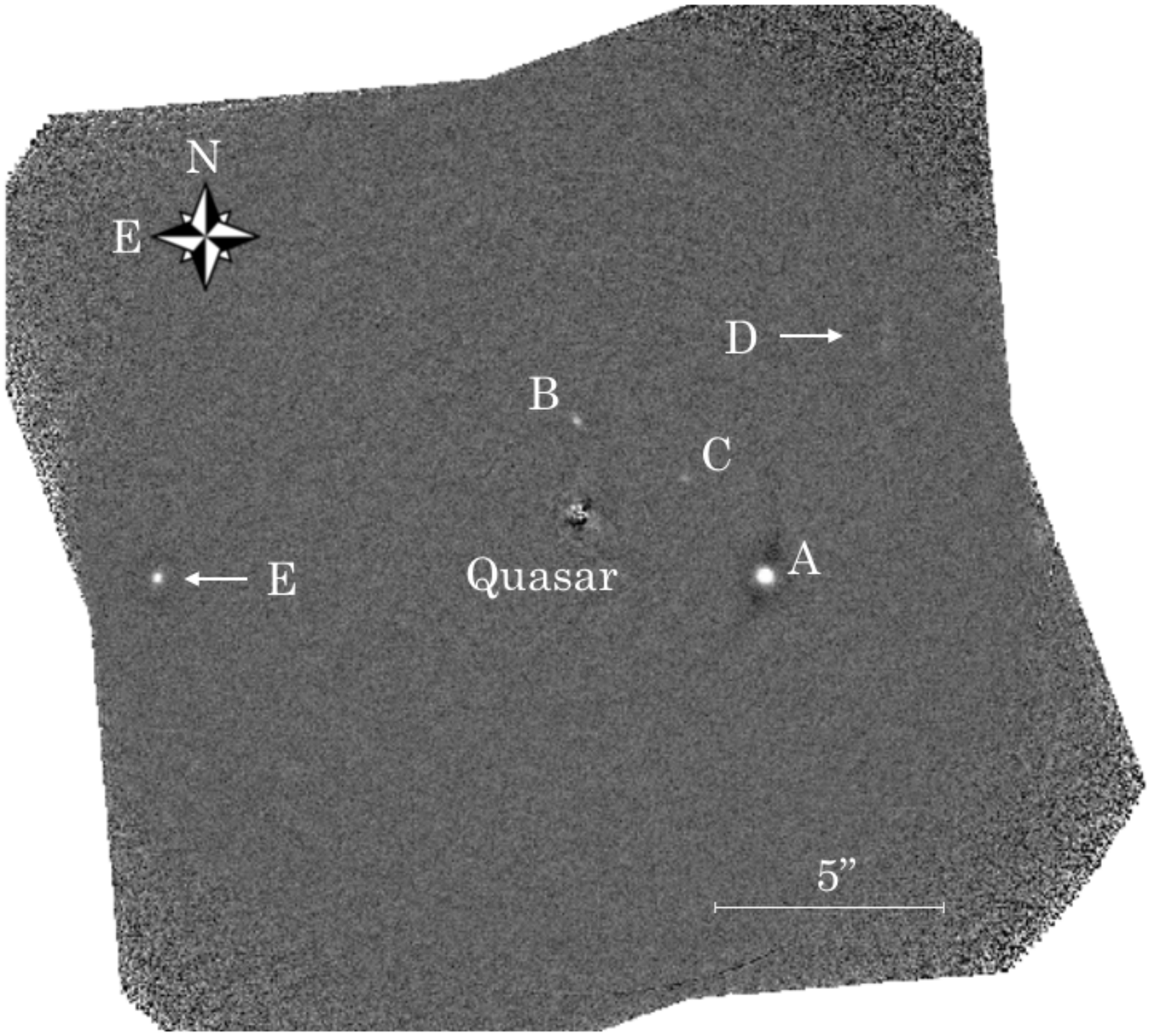} \\
    \centering\small (c) Final Image with PSF subtraction
  \end{tabular}
  \caption{Quasar PSF (a), and final reduced image without (b) and with (c) PSF subtraction. Objects considered for this study are marked A, B and C.  Two other new objects (D and E) at higher impact parameters are indicated with arrows, but are not considered further in this paper.  The central object labeled 'Quasar' in frame (c) is the residual image of the quasar after PSF subtraction. The dark arc around the brightest object in frame (c) is caused by the residual signature from the median of the object as it moved in PA.  The faint diagonal line slightly below the quasar in frame (b) is from a detector artifact that was not well subtracted by the dark frame.  No large scale structure is seen around objects A, B or C.}
  \label{fig1}
\end{figure}

The final reduced and combined NIRI image is shown in Figure~\ref{fig1}, in which 5 objects (A, B, C, D and E) are detected with angular separations from the QSO ranging from 2 to 10 arcsec.  Based on previous DLA host galaxy studies, impact parameters are typically a few tens of kpc with higher log N(HI) systems being systematically closer to the QSO \citep[e.g.][]{moller98,chen,cooke,monier}.  \citet{rao} find a median impact parameter for DLA hosts of 17.4 kpc and \citet{reeves} note that DLA discoveries with impact parameters greater than 20 kpc are 'extremely rare'.  Larger separations have been concluded to more likely be associated with multiple absorbers \citep{ellison07}.  On the basis of these studies, and the high log N(HI) of QSO J1431+3952,  an upper limit of $b$ = 30 kpc, or $\sim$ 5 arcsec at a redshift of $z=0.6$, will be used to distinguish candidate host galaxies from field interlopers in this study.  Three of the five objects in our reduced image (A, B and C) fulfill this impact parameter criterion.  Object D is located to the northwest of the quasar at 8.1$\pm$0.1" (55.8$\pm$0.7 kpc) and object E to the east (9.70$\pm$0.07", 66.8$\pm$0.5 kpc). Given their large impact parameters, D and E are excluded from further study in this work.   

\begin{figure}[H]
\epsscale{.81}
\plotone{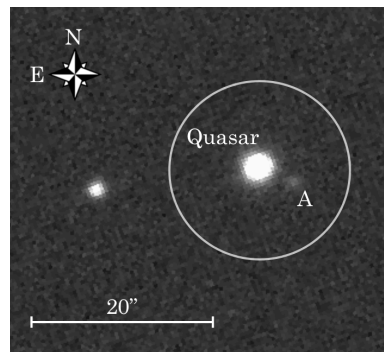}
\caption{Section from an SDSS Image, field 391, run 3813, camcol 3, i-band.  The NIRI field-of-view visible for all angles of rotation is indicated by the white circle, which is centred on the quasar. The bright object east of the quasar is not visible in our field of view.  Object A in this figure is the same object as the one labelled "A" in Figure~\ref{fig1}.   North is up and East is left. }
\label{sdss}
\end{figure}

The brightest of the three objects considered in this study, A (at an impact parameter of $\sim$ 30 kpc), was faintly visible before the raw images had been processed and is also visible in archival SDSS imaging, which we show for comparison in Figure~\ref{sdss}.  This is the same galaxy identified by Ellison et al. (2012) and Zwaan et al. (2015).  However, as mentioned above, the photometric redshift of galaxy A ($z=0.08$) is inconsistent with that of the DLA ($z_{abs}=0.602$).  Galaxies B and C (at impact parameters $\sim$ 15 and 17 kpc respectively) are identified in our NIRI image for the first time.  We return to the discussion of the most likely absorber in Section~\ref{disc}.  The darker arc around object A is due to a residual signature from the median of the object as it is moved in position angle and the dark circular halo is due to the unsharp mask.


\subsection{Candidate Galaxy Simulation and host parameters}\label{simul}

Having made basic identifications of three candidate absorbing galaxies within our impact parameter search radius, more precise measurements of their magnitudes, locations and FWHM were determined through forward modelling.  This technique involved creating replicate objects with varying parameters, inserting them in the individual images and then determining which of the simulated objects best matched the real objects after going through image subtraction and combination.  The advantage of this method is that it accounts for any missing flux that is removed with combining the images and PSF subtraction.  For example, if an extended object overlaps in adjacent images, some flux would be lost in the final image, and negative wings would be apparent.  However, if a model was created for each image, not just the combined, it would need to exactly match the actual object to precisely reproduce the final image.

First, the initial parameters of the objects were determined by fitting a 2D gaussian to the three candidate galaxies in the final reduced image.  Simulated images were then created by placing 2D gaussian objects created from the initial parameters at the same location as the candidate galaxies in each exposure in an empty frame with the same dimensions as the images.  In order to account for nonlinear rotation during the 60 second exposure, instead of placing one gaussian at the object's location, five gaussians were placed to match the rotation of the parallactic angle, as five was sufficient to span the spread in parallactic angle.  This made the model slightly smeared along the direction of rotation.  While this wasn't necessary for the modelling of the fainter objects, B and C, at low impact parameters, it was critical for the brightest object, A, which was at the highest impact parameter.  These simulated gaussians, whose total flux was normalized to one, were convolved with the PSF of the quasar which was taken from the inner one square arcescond around the quasar PSF image (see Figure~\ref{fig1}a).  This PSF image, instead of just a core PSF model, was used to avoid loosing flux from the PSF wings so that the derived object flux and magnitude can be properly calculated. Each simulated frame was then processed in the same manner as the real exposures were - PSF subtraction over the entire image, rotate north up, median combine all exposures from each night and combine nights with a SNR weight.  Each object was multiplied by a scaling factor to best match its intensity to its pair's real intensity.  The simulated image was subtracted from the original image.  If the objects had been modelled perfectly, there would be no residual image left and the noise would match the background level.  To test this, the standard deviation in a 1.5$\lambda$/d aperture was found at the location of the object and compared to the standard deviation in the noise at a similar sized annulus around the aperture.  Each object's location, FWHM, and scaling factor was adjusted until the noise was minimized in the aperture.  For each object, optimization of the parameters reached the background noise level.  The parameters from the models that minimized the noise were then taken to calculate the magnitudes and locations for each object.   Position angles (PA) were calculated counterclockwise from the North axis. 

\begin{figure}[H]
\epsscale{.85}
\plotone{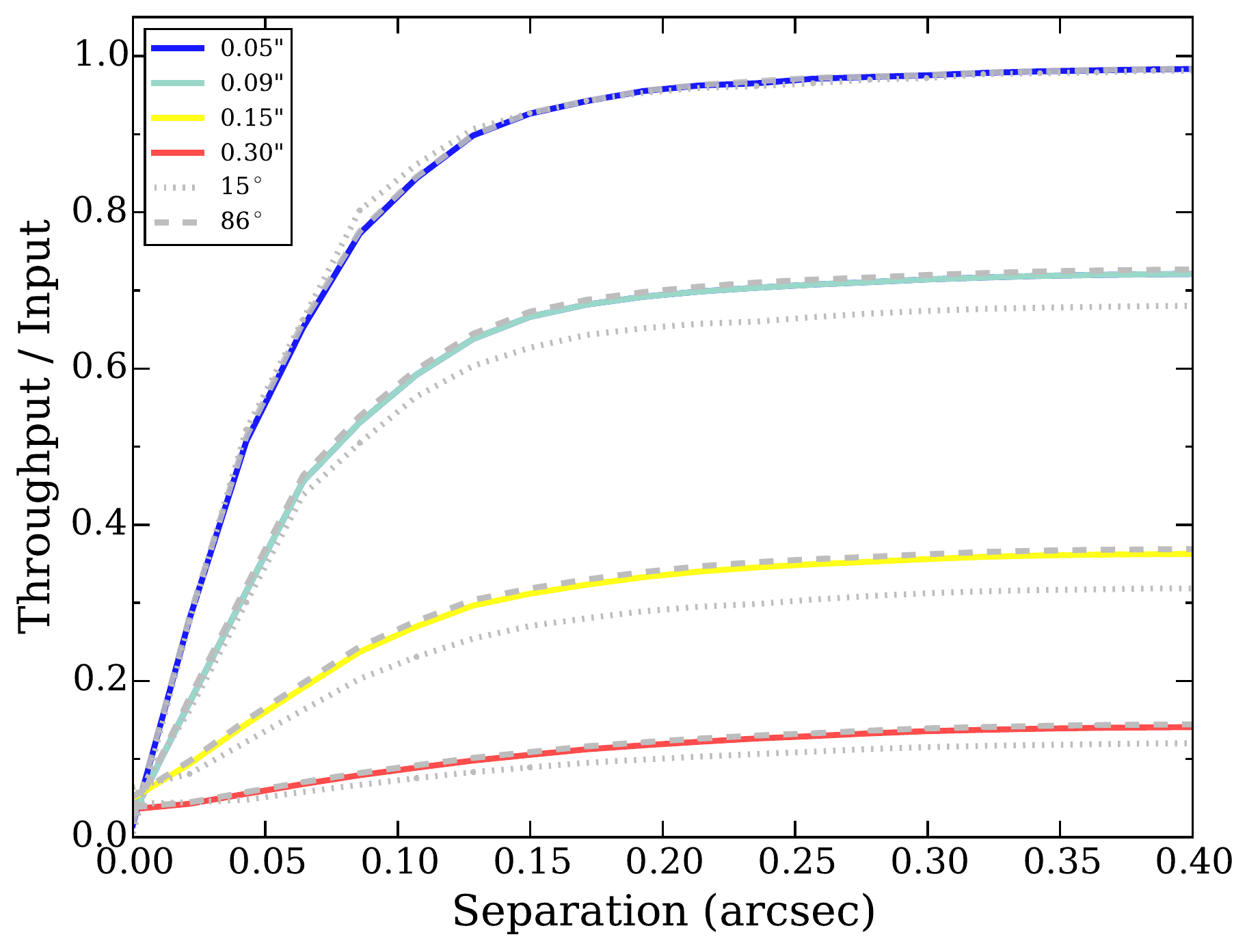}
\caption{Throughput as a function as radius and model object size for both nights.   Grey dashed and dotted lines indicate the separate throughput for each night (different FOV rotation) and the coloured line shows the combined throughput.   The throughput varies sightly for the different field-of-view rotations. The throughput only drops below 90\% for the inner 0.13 arcsec (0.88 kpc) when the model FWHM is 0.05 arcsec.  For larger models, the throughput drops significantly with a maximum of 37\% for a 0.15" FWHM object.}
\label{throughput}
\end{figure}

Since throughput can decrease at low impact parameters, simply measuring the flux at small annuli can be misleading.  To adjust for the reduction in throughput due to partial self-subtraction, a calibration was performed by inserting artificial unresolved (FWHM of 0.05") model galaxies (gaussians normalized to one) at different separations from the centre.  These models were inserted in the same manner as described above, except that instead of placing the models in the same locations as the objects, they were placed at the same PA with different separations.  Measuring the models' peak flux in the final image, as opposed to the peak flux of the original inserted model gives the percentage of throughput.  The throughput was measured for each night separately as the nights had different field-of-view rotations, which effects the throughput.  The throughput as a function of radius and model object size can be seen in Figure~\ref{throughput}.  A polynomial was fit to the throughput percentage as a function of separation and was used to adjust adjust each pixel in the original image based on the pixel's separation from centre of the image.  These adjustments were applied to the stacked image from each night before they were combined together.  Once each night had been adjusted, they were combined with a weighted mean and the dispersion was calculated at each annulus.  The throughput corrections only affected the inner 1 kpc of the image when correcting with a 0.05" FWHM model galaxy.  For the final images, a throughput correction using a 0.05" model was applied, although others were tested (Figure~\ref{throughput}).  The unsharp mask removes the low spatial frequencies of the larger, diffuse models, significantly reducing the flux throughput.  However, the simulated models for the throughput are smooth gaussians, they are not perfect galaxy analogues; galaxies typically have a higher concentration of light at the core. Thus, these throughput models represent the worse-case-scenario of an extremely diffuse galaxy with no central bulge.  Any real galaxy likely has a more compact profile, and even more structure, like spiral arms, which would increase the throughput and detectability. A more detailed study of throughput using real galaxy images is beyond the scope of this study.


To determine the uncertainties associated with the astrometric measurements of location, FWHM and the intensity scaling factor from the simulated images, 10 model templates were created in the same manner as described above, except instead of placing the models at the same position angles as the objects, they were placed at different position angles and the separations of the original galaxies.  Again, exposures and nights were combined and scaled in the same manner and subtracted from the final image.  The locations and FWHM for each of the 10 images were recovered and measured, and their standard deviation was taken as the error.  For the magnitude uncertainty, an additional intensity scale model (at the same scaled intensity of the model that matched the object) was subtracted from the models at different PA.  The magnitude uncertainty from modelling was added in quadrature with the uncertainty of the quasar's magnitude for the total magnitude uncertainty.  

The three candidate host galaxies (A, B and C) were characterized using the measurements from the simulated images.  To find the galaxies' magnitudes, the simulated image was flux calibrated using an image combined without the PSF subtraction and the quasar magnitude from a photometric catalogue \citep{schneider}. The apparent K-band magnitudes of the galaxies A, B and C were 16.74$\pm$0.06, 19.8$\pm$0.1 and 20.4$\pm$0.2 respectively.  The impact parameters of the galaxies are  30.370$\pm$0.004, 14.7$\pm$0.4, 17.3$\pm$0.3 kpc, assuming the redshift of the DLA.  A full list of the model-derived parameters can be found in Table~\ref{galxprops}.

The FWHM of the gaussian modelled to each object was used to determine the angular size of the object.  By taking the FWHM found through forward modelling, we have allowed the FWHM to be a free parameter and thus are able to empirically derive the object size.  The FWHM of the quasar was 0.050" $\times$ 0.056", and is taken as the final angular resolution achieved with AO, consistent with a diffraction limited PSF produced by the AO system (theoretical value of 0.067").  Objects B and C were larger (0.058$\pm$0.006" $\times$ 0.079$\pm$0.008" and 0.075$\pm$0.007" $\times$ 0.057$\pm$0.006") than the angular resolution but only in one direction, which suggests these objects may be elongated. Object A  was significantly smaller (0.01$\pm$0.03" $\times$ 0.01$\pm$0.03") than the resolution and thus is clearly unresolved.  The FWHM sizes of objects B and C  correspond to $\sim$$0.3-0.5$kpc.  Although these sizes seem quite small for a galaxy at the redshift of the absorber, it is likely this measurement corresponds to only the bulge of the galaxy which is bright enough for our detection; the actual galaxy could be much larger.

\subsection{Mass Calculations}\label{mass}

The K-band magnitudes of the galaxies were converted to approximate stellar masses using the luminosity to stellar mass relation detailed in \citet{long}.  The conversion derived in this paper is to deal specifically with early type galaxies and assumes a formation redshift of 4.  To convert the magnitudes into masses for a certain wavelength, $\lambda$, the following equation is used:

\begin{equation}
\begin{aligned}
log(M_{gal}) = log(M/L_\lambda)+0.4kcor_\lambda +2 log(d_{pc}) \\
	- 2.0+0.4M^\sun_\lambda - 0.4m_\lambda ,
\label{eq:mass}
\end{aligned}
\end{equation}

where $M/L_\lambda$ is the mass to light ratio solar units, $kcor_\lambda$ is the k-correction, $d_{pc}$ is the distance in parsecs, $M^\sun_\lambda$ is the absolute magnitude of the Sun, and $m_\lambda$ is the apparent magnitude of the galaxy.  $kcor_\lambda$ is calculated with the absorber redshift and parameters detailed in \citet{long}.  For the K-band,  $M^\sun_\lambda$ is 3.41 \citep{allen}.  The mass to light ratio assumes the redshift  ($z$) derived from the DLA spectra and various initial mass functions (IMF):

\begin{equation}
(M/L_\lambda)=a_0 +a_1z +a_2z^2 +a_3z^3 ,
\label{eq:ml}
\end{equation}

where the coefficients $a_i$ are parameterized for six different IMF models.  For a Chabrier IMF and  \citet{bruzual} models, the coefficients in order are: 0.03, 0.10, -0.008, 0.0004. The Chabrier model based on code from \citet{bruzual} is used throughout this paper for comparison with models from \citet{christensen} and \citet{maiolino}.  Log stellar masses of 11.1, 9.9 and 9.7 M$_\sun$ are calculated for objects A, B and C respectively.  Systematic uncertainties in mass-luminosity relationships due to stellar population synthesis (SPS) and IMF models can be hard to quantify.  
 \citet{long} derive their mass retrievals for their mass estimators using two mock galaxy catalogues.  For the K-band mass estimates used in this paper, \citet{long} find 1$\sigma$ errors of $\sim$$30\%$ from their mock galaxy catalogues. These errors are similar though slightly larger to the results of other studies, perhaps due to \citeauthor{long}'s (\citeyear{long}) use of only one band whereas other studies looked at multiband SED fitting.  We adopt a conservative $30\%$ systematic error which is added in quadrature to the photometric error in calculating the final mass with uncertainty.  A list of the masses and their uncertainties for each object can be found in Table~\ref{galxprops}. 
 
\begin{figure}[H]
\epsscale{1.15}
\plotone{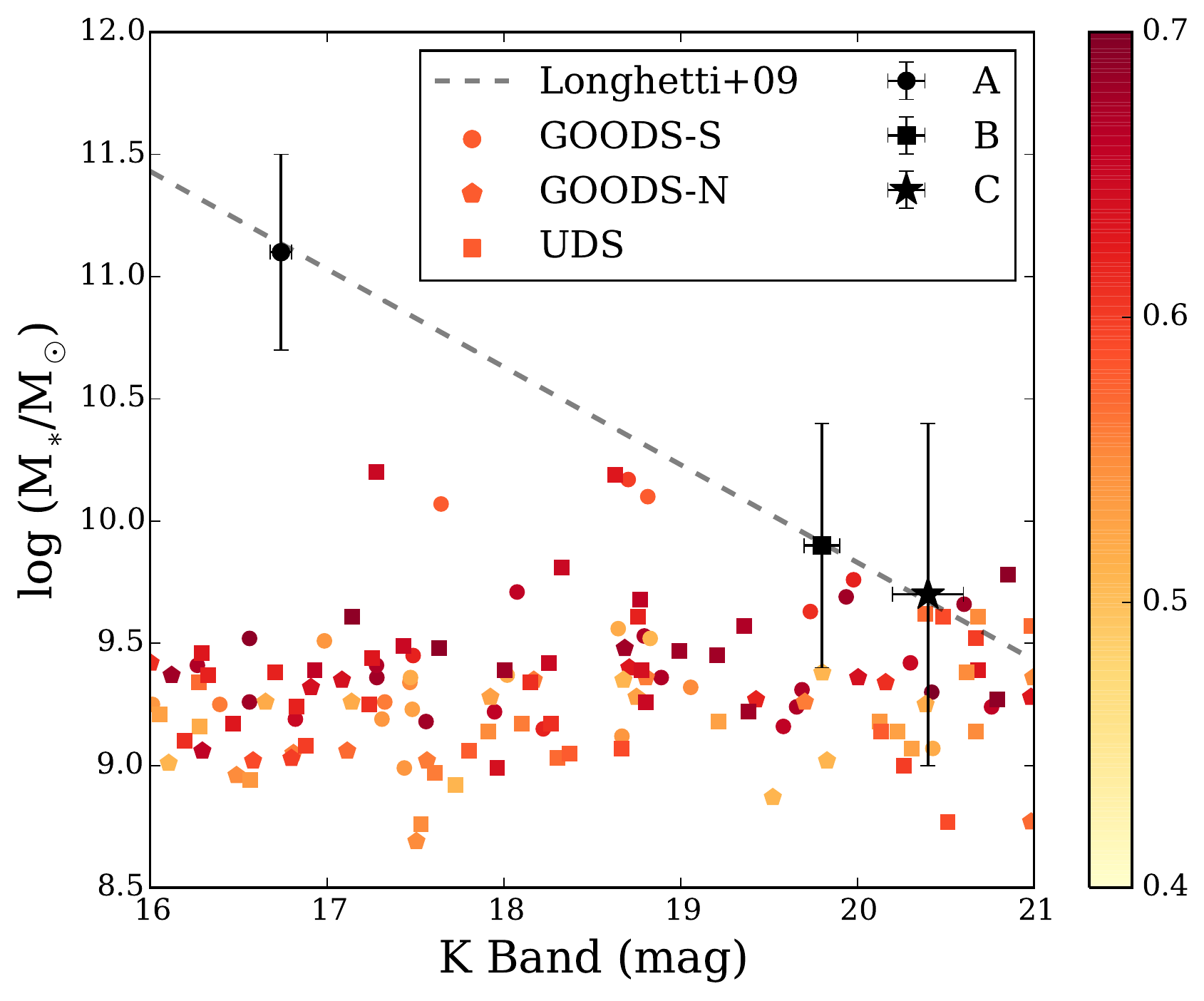}
\caption{Objects from this study shown with galaxies from the 3D-HST catalogue.  Galaxy colours indicate the redshift of the galaxy.  Shapes are indicative of the different fields from the catalogue. }
\label{catcompare}
\end{figure}

To check if the assumptions under the \citet{long} were valid, the masses and magnitudes obtained for each object were checked with galaxies in the GOODS-S, GOODS-N and UDS fields in the 3D-HST catalogue \citep{skelton,brammer}.  Galaxy masses in this catalogue are calculated with the FAST code \citep{kriek} using the \citet{bruzual} stellar population synthesis model library with a Chabrier IMF at solar metallicity.  Spectroscopic redshifts were used when available, and photometric redshifts when spectroscopic data was unavailable for dim objects.  Compared with the \citet{long} estimates, the 3D-HST catalogue shows lower masses at brighter magnitudes (see Figure~\ref{catcompare}).  However, objects B and C are consistent with the 3D-HST galaxies, suggesting that the mass estimate is within reason.

Our derived sizes and masses can be compared to known galaxy scaling relations.  In the local universe, galaxies with masses similar to objects B and C (log(M$_*$) = 9.9 and 9.7  M$_\sun$ respectively) are found to have effective radii of 1 -- 1.5 kpc, in contrast to the measured FWHM values of 0.3 -- 0.5 kpc.  However, at higher redshifts, galaxies tend to be more compact, with sizes up to a factor of two smaller at $0.5<$ z $<1$ than in the local universe \citep[][and references therein]{fan}.  The sizes measured for our galaxies are therefore plausibly in the range expected at these redshifts.  Indeed, in \citet{van} there are several cases of $\sim$ 10$^{10}$ M$_\sun$ galaxies with sizes less than 0.5 kpc at intermediate redshifts.  We conclude that the sizes derived from our observations are not in conflict with galaxy size-mass relations at $z \sim 0.6$, supporting their candidacy as the absorbing galaxy.

\subsection{Sensitivity Curves}\label{curves}

With formulations connecting K-band magnitudes and stellar mass, it is possible to assess the sensitivity of our observations to detections as a function of M$_{\star}$ and impact parameter. The detection limits were found by calculating the dispersion in a 1.5$\lambda$/D width annulus using the IDL \texttt{robust\_sigma.pro} routine using the final image that had been throughput corrected (see Section ~\ref{simul}) so as to accurately reflect the sensitivity at low impact parameters. The dispersion was flux normalized to the quasar's flux and converted into apparent magnitude.  A five sigma detection level, or five times the dispersion of the background flux, is assumed to ensure detection. 

\begin{figure}[H]
\epsscale{.95}
\plotone{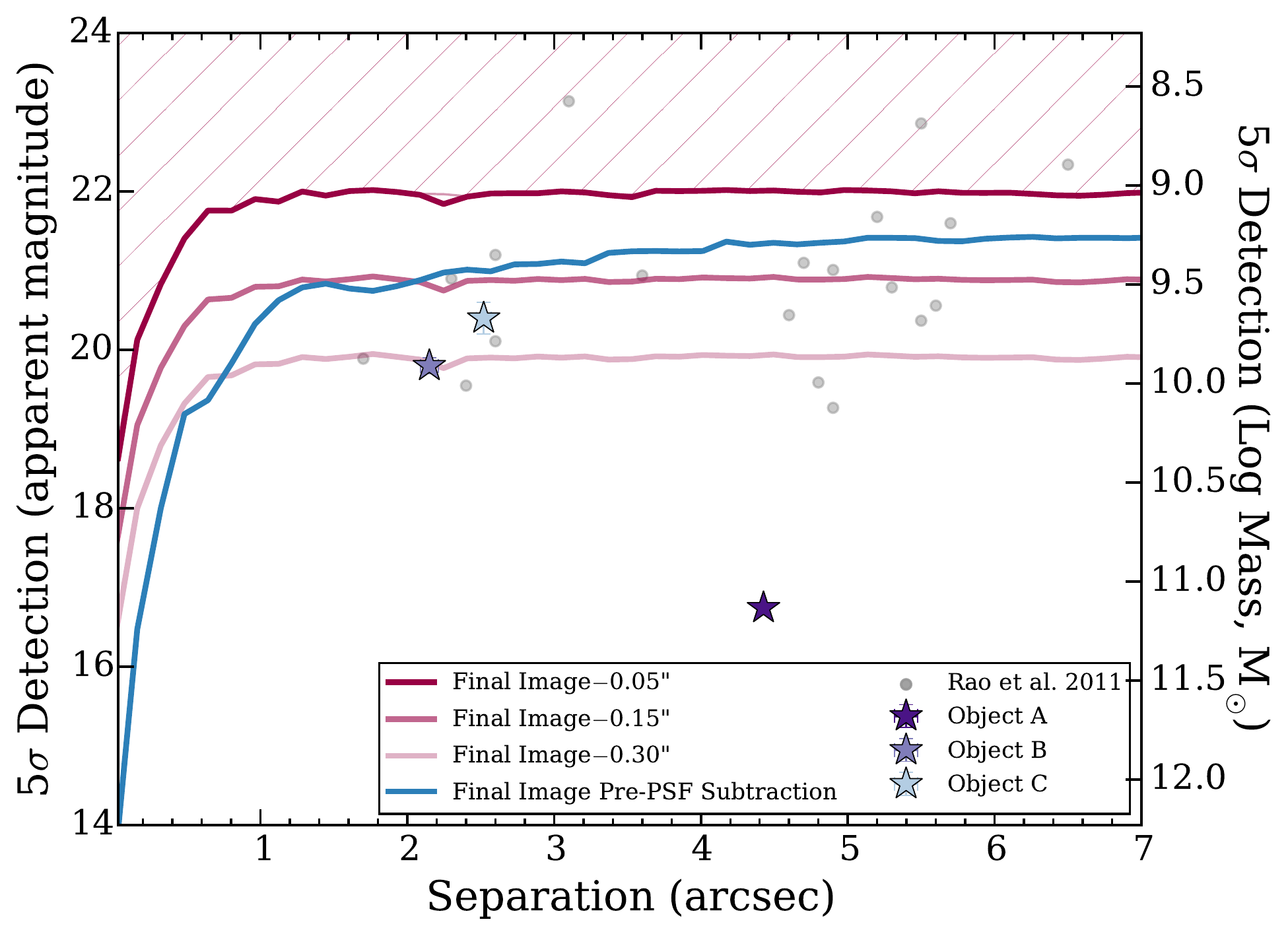}
\caption{Parameter space sensitivity for DLA host detection using the methods presented in this paper.    The red curve indicates the detection limits of this study, such that objects with combinations of K-band magnitude and impact parameter in the hashed region could not have been detected.  The three red lines indicate the detection limits after correcting throughput for different model galaxy FWHM sizes.  The blue curve shows the detection limit for the final image that has not been PSF subtracted and with a throughput correction of a 0.05" model galaxy.  The quasar's PSF limits the detection level, particularly at low impact parameters.  Masses on the right hand y-axis were converted from the apparent magnitudes on the left hand y-axis using the method of \citet{long} described in Section~\ref{mass}.   The three detected galaxies within our impact parameter threshold are shown at their measured magnitudes (and stellar mass equivalent) and separations as star symbols.  All magnitudes are in the K-band; photometric errors are typically smaller that the sympbol.    For comparison, we also show previously identified DLA host galaxies compiled in \citet{rao}.    }
\label{contrast}
\end{figure}

The sensitivity curve is shown in Figure~\ref{contrast}, were the hashed region shows the combination of K-band magnitude and impact parameter of galaxies that could not have been detected in our final image.  For images that have not been PSF subtracted, the limit for detection is noticeably shallower, especially at low impact parameters, i.e. the PSF subtraction dramatically helps increasing detectability at low impact parameters where DLA hosts are most likely to be.  The PSF subtraction additionally helps at larger radii where it acts as a background sky and detector artifact subtraction, which is why the PSF subtracted sensitivity curve never overlaps the non-PSF subtracted curve.  The three galaxies (A, B and C) within our impact parameter threshold are shown as star symbols, where the errors are typically smaller than the symbol size.    We note that the range of M$_{\star}$ and impact parameter space in which candidates could have been detected extends to even lower values than those exhibited by the galaxy candidates detected in this study.  For example, we could have detected a $10^{9.2}$ M$_{\odot}$ galaxy (approximately 1/10 the mass of the Milky Way) as close as 0.5 arcsec (3.4 kpc).  At impact parameters beyond $\sim$ 6 kpc, our mass sensitivity plateaus at $\sim$ log $10^{9.0}$ M$_{\odot}$.  However, a caveat to this method of deriving sensitivity curves is that it is more accurate for galaxies that are close to being unresolved point sources, as extended galaxies with more diffuse emission are harder to detect after PSF subtraction.  This is shown with the three red curves which use model galaxies of 0.05" to 0.30" FWHM to correct for throughput (see Section~\ref{simul}).  For throughput correction with a larger (0.30"), diffuse model galaxy, the detection limit decreases by an of magnitude.  Although this may seem significant, this is a worse-case scenario as most galaxies will not be entirely diffuse and would still have a visible dense core or spiral arms.  For comparison, Figure~\ref{contrast} also shows the positions of DLA host galaxies detected in \citet{rao} for comparison.  Host galaxies were only selected from \citet{rao} if they had K-band magnitudes, b $\textless$ 10 arcsec, and have been confidently identified using spectroscopic redshift, photometric redshift or colours. 

\section{DISCUSSION}\label{disc}

Until now, the only object previously identified within 50 kpc of the J1431+3952 quasar sightline was object A  \citep{ellison12,zwaan}, see Figure \ref{sdss}.  Since it has been previously proposed  \citep{kanekar}  that DLAs with low spin temperatures and high metallicities such as this one are associated with relatively luminous disk galaxies, object A, which we have estimated in this work to have a stellar mass log (M$_{\star}$/M$_{\sun}$) $\sim$ 11 might initially seem like a compelling candidate.  However, its low photometric redshift \citep[$z=0.08$,][]{ellison12} is inconsistent with the DLA redshift ($z_{abs}=0.602$){\footnote{Results in \citet{abazajian} indicate that SDSS galaxies assigned a photometric redshift of 0.1 never have spectroscopic redshifts above 0.3.}.  The photometric redshift was based on limited SDSS photometry, so it is nonetheless interesting to further explore the possibility of A as the DLA host. With the identification of additional candidates B and C, we can further explore the likelihood of each as the absorbing galaxy, by considering scaling relations of various physical properties. 

We begin by exploring the trend between impact parameter and neutral hydrogen column density of DLA hosts.  This trend has previously been noted by several studies which find an inverse correlation between the neutral hydrogen column density and impact parameter \citep[e.g.][]{krogager,moller98,peroux,monier, rao,christensen07}.  For example, \citet{rao} report the anticorrelation at a 3$\sigma$ level significance.  Figure~\ref{nhi-b} shows the results from our study combined with previously identified and confirmed  DLA hosts  \citep{chun10,peroux,krogager,fum}.  Objects B and C are consistent with the general anti-correlation seen in previous studies, whereas object A is at a significantly higher impact parameter than would be expected for its N(HI), even given the high spread in the trend.   In addition to its inconsistent photometric redshift, Figure~\ref{nhi-b} therefore adds further evidence against object A being the DLA host.

\begin{figure}[b]
\epsscale{.95}
\plotone{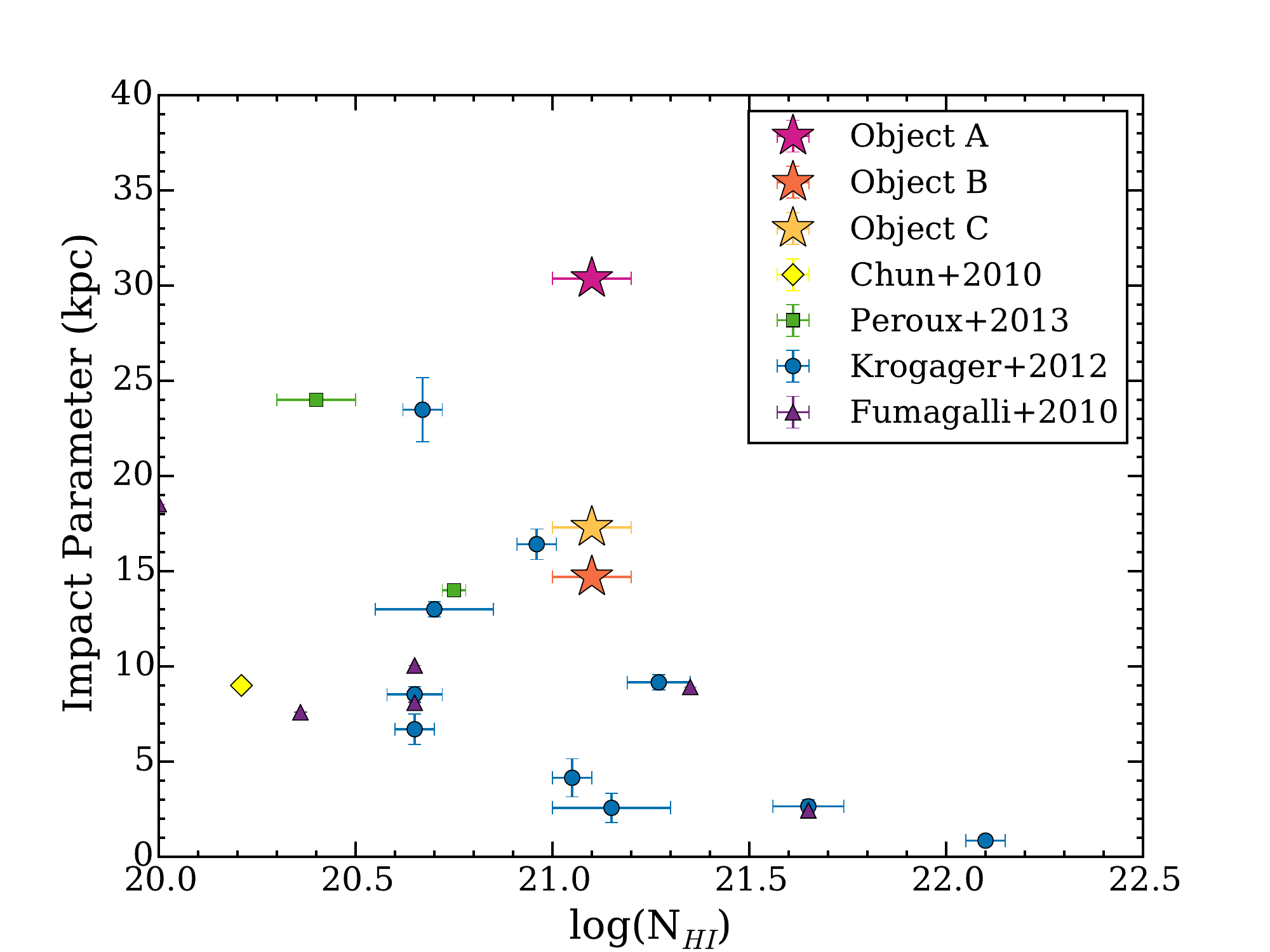}
\caption{Impact parameter plotted with log HI column density for the three objects in this study as well as four other studies \citep{chun10,peroux13,krogager,fum}}
\label{nhi-b}
\end{figure}

Luminosity or stellar mass-metallicity (MZR) relations have been well studied across a range of redshifts starting with \citeauthor{lequeux} in 1979. Based on the high $z$ mass-metallicity relation described in \citet{maiolino}, \citet{christensen} established the relationship for DLAs at a range of redshifts and derived a correction for the impact parameter.  Specifically,  \citet{christensen} add 0.022$b$ (where $b$ is the impact parameter in kpc) to the absorption line metallicity to account for the impact parameter.  Figure~\ref{mzr-mh} shows the three candidate DLA galaxies from our study compared with the mass-metallicity relation from \citet{maiolino}.  We adopt the common approach of using zinc as our elemental tracer of metallicity, using the metallicity reported in \citet{ellison12}.    Without any correction for impact parameter, the absorption line metallicity of the DLA combined with the stellar masses of the 3 candidate host galaxies identified in this study all fall below the mass-metallicity relation presented in \citet{maiolino}, as shown by open stars in  Figure~\ref{mzr-mh}.    After the application of the impact parameter correction to the metallicity, all of the candidate host galaxies move closer to the expected MZR, although none are a particularly good match. \citet{christensen} find a scatter in their relation of 0.39 dex in log $M^{DLA}_{*}$ which encapsulates only object C.

\begin{figure}[h]
\epsscale{.9}
\plotone{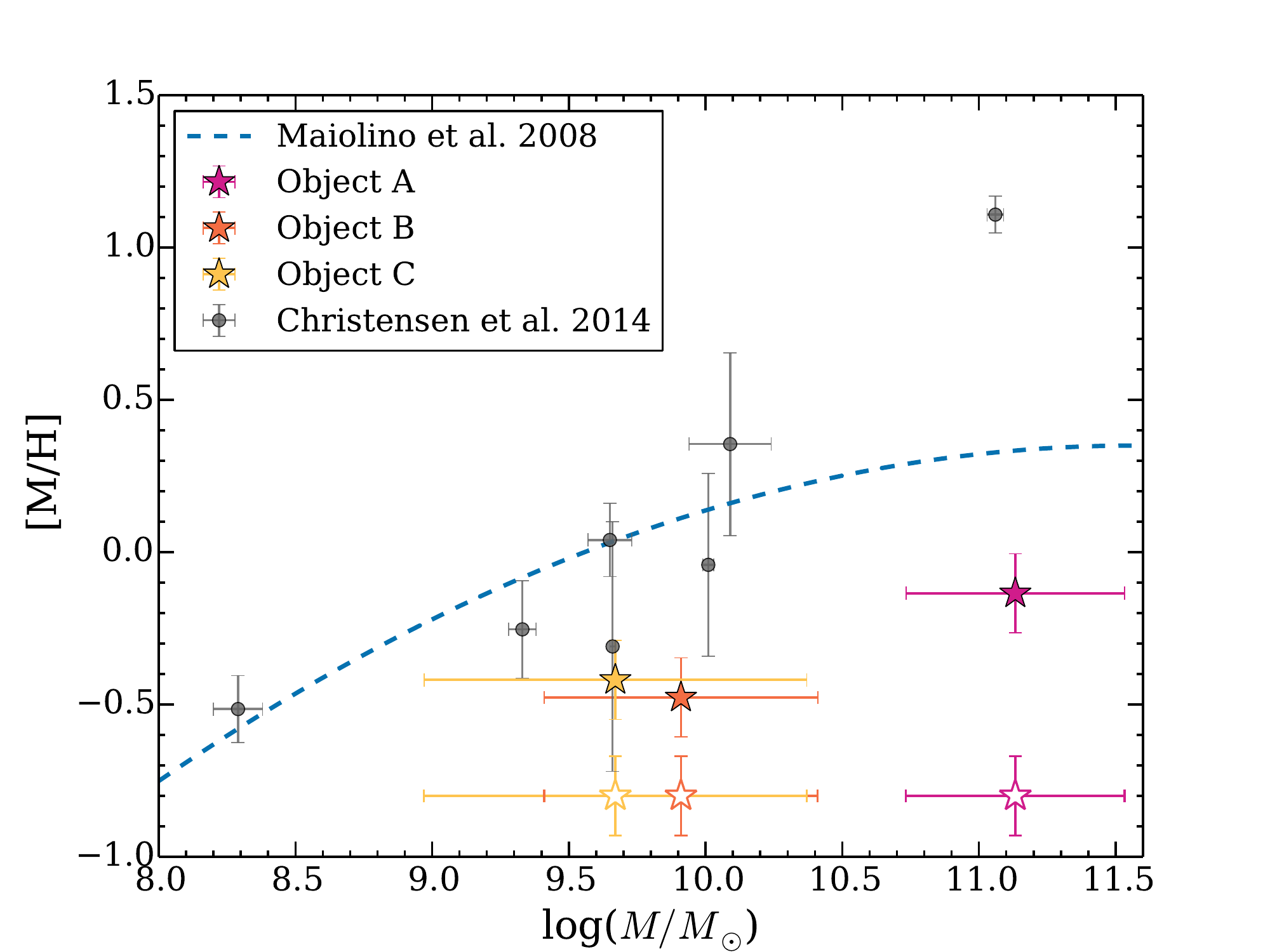}
\caption{MZR fits from \citet{maiolino} plotted with the objects from this study (open stars).  Once impact parameter adjustments from \citet{christensen} have been applied (filled stars), the objects are closer to the established MZR.  Zinc measurements from \citet{ellison12} are used as a proxy for metallicity for the three objects.  DLAs from \citet{christensen} with z $<$ 1 (median z=0.52) are shown as grey circles for comparison.}
\label{mzr-mh}
\end{figure}

An alternative way to use the MZR relation is to assume that the host galaxy follows the relation derived by Christensen et al. (2014), and use the absorption line metallicity to determine the predicted stellar mass as a function of impact parameter.  This calculation is shown in Figure ~\ref{contrast2} as a dashed line, over-plotted with our sensitivity curve.    Figure ~\ref{contrast2} shows that \textit{if} the host galaxy follows the impact-parameter dependent MZR derived by Christensen et al. (2014), it would have been detectable only if log M$_{\star}$/M$_{\odot} >$ 9, with corresponding impact parameter $>$ 30 kpc.  Although we do have two further detections in the NIRI field at larger impact parameters (objects D and E), we have shown in Figure \ref{nhi-b} that DLAs with column densities as high as the sightline studied here (log N(HI)=21.1) are rarely found at high impact parameters.  

Although the dashed line in Figure \ref{contrast2} may seem to imply that our study is not particularly sensitive to galaxies on the MZR for this DLA, significant scatter may be expected.  We therefore use the sample of DLAs  in the redshift range $z=0.3-0.9$ with zinc detections (for metallicity comparison) from \citet{berg} to gauge the possible scatter around Christensen et al. (2014)'s best fit relation.  The result is shown in the gray shaded region in Figure \ref{contrast2}, in which it can be seen that host galaxies with sufficiently high metallicities following the MZR would be detectable to separations as low as 1.7 kpc for true point source galaxies.  Moreover, galaxies B and C lie inside this shaded region, implying that they are plausible candidates for the DLA host.  

The sensitivity limits shown here were derived for a 16th magnitude (R-band) quasar observed with an 8-meter telescope.  Observing fainter quasars would increase detection limits in the region limited by quasar PSF residual (i.e. the inner one arcsecond).  However, the current limits of AO require a 17.5 magnitude quasar or nearby star for tip-tilt corrections.  The rest of the image outside of one arcsecond, which is background noise limited, could be improved with longer integrations, as contrast increases with the square root of integration time.  As the next generation of thirty-meter telescopes become available, larger apertures will increase resolution and allow for quasars 16 times fainter to be observed with AO, allowing for detections of host DLA at 3 times lower impact parameters and 16 times fainter (3 mag deeper contrast) in the background-limited regime.  

\begin{figure}[H]
\epsscale{.85}
\plotone{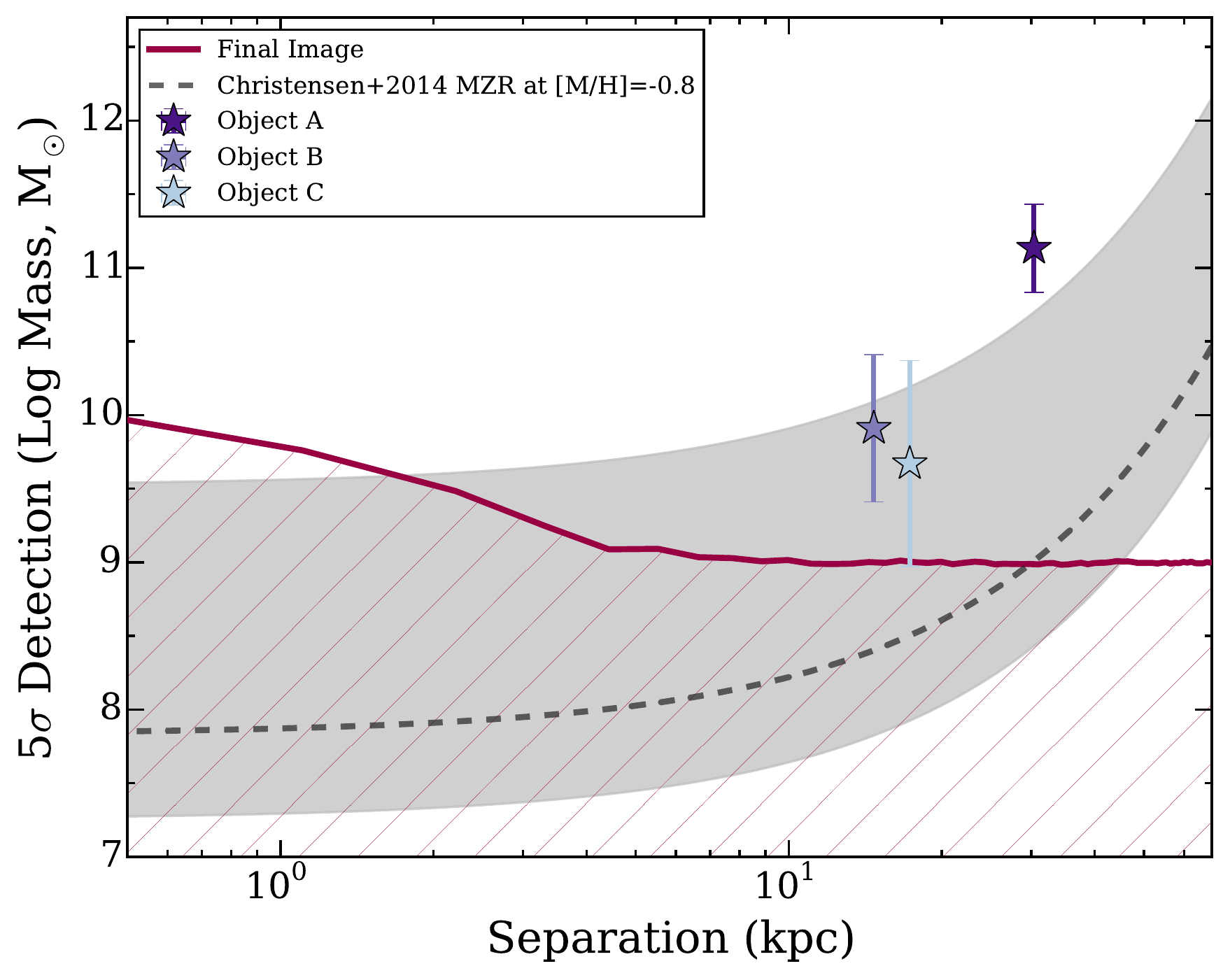}
\caption{If we look at the sensitivity curves as a function of physical separation, we can see that assuming the DLA host follows the \citet{christensen} MZR with the redshift and metallicity fixed to the DLA (grey dashed line), the host would only have been detected in this study if its stellar mass log (M$_{\star}$/M$_\sun$) was greater than 9 and separation greater than 30 kpc.  The grey shaded region shows a range of MZR for other known DLA hosts in the redshift range $z=0.3-0.9$ \citep[DLA catalogue from][]{berg}.  DLA hosts following the MZR could be detected with the techniques described in this paper at impact parameters low as 1.7 kpc. For the redshift range selected, changes in the conversion of arcsec to kpc is negligible for the sensitivity curves.}
\label{contrast2}
\end{figure}

\begin{table*}
\centering
\caption{Candidate Galaxy Characteristics}
\begin{tabular}{p{4cm}p{2cm}p{2cm}p{2cm}}
    \toprule
    \midrule
    \bfseries Object & 
    \bfseries A & 
    \bfseries B & 
    \bfseries C \\ 
    \midrule
    Image  & \adjustimage{height=2cm,valign=m}{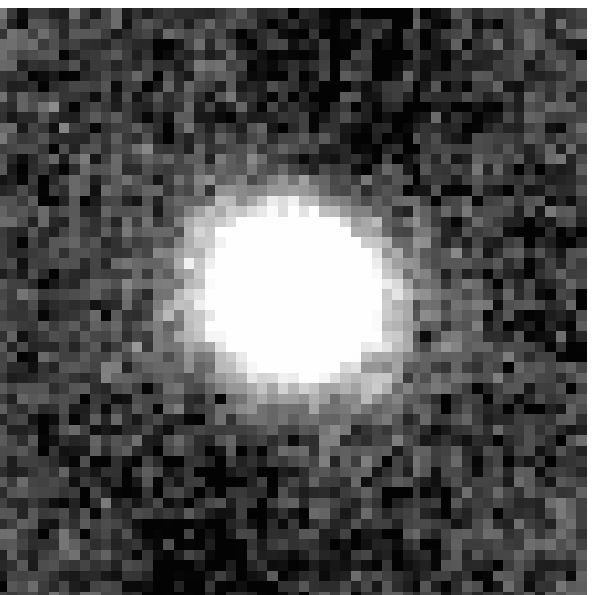}& \adjustimage{height=2cm,valign=m}{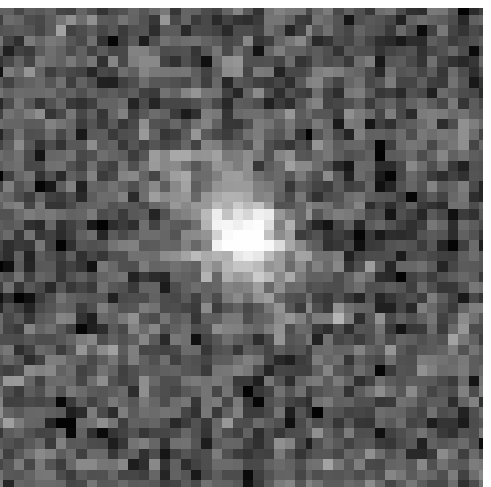} & \adjustimage{height=2cm,valign=m}{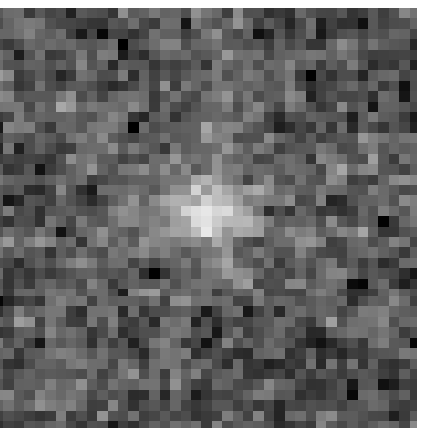}\\
  Magnitude (K band) & 16.74$\pm$0.06 & 19.8$\pm$0.1& 20.4$\pm$0.2 \\
Impact Parameter (arcsec) & 4.4260$\pm$0.0006 & 2.14$\pm$0.06 & 2.52$\pm$0.05\\
Impact Parameter (kpc) & 	30.370$\pm$0.004	 & 14.7$\pm$0.4 &17.3$\pm$0.3	\\
Position Angle (deg) &251.701$\pm$0.009 & 2.61$\pm$0.08 & 289.32$\pm$0.07 \\
Log Mass (M$_\sun$)& 11.1$\pm$0.4 & 9.9$\pm$0.5 & 9.7$\pm$0.7\\ 
    \midrule
    \toprule
   \tablecaption{Candidate Galaxy Characteristics}
\end{tabular} \\
NOTE. --- Impact parameter in kpc calculated assuming the redshift of the absorber.
\label{galxprops}
\end{table*}

From comparisons to scaling relations between impact parameter and N(HI), and stellar mass with metallicity, we have shown that galaxies B and C (log M$_{\star}$/M$_{\odot} =$ 9.9, 9.7 respectively) are plausible candidates for the host galaxy of the DLA towards J1431+3952.  Alternatively, the host could be an as yet unidentified galaxy with relatively low stellar mass, with the MZR constraining the limit to be log M$_{\star}$/M$_{\odot} <$ 9, or it could be located at a very small impact parameter (almost inline), making it making it impossible to see in these reductions.  In none of these scenarios is the host galaxy particularly massive, falling several times below the mass of the Milky Way.  Kanekar \& Chengalur (2003) noted that low T$_s$ DLAs tend to be associated with luminous ($l \sim$ L$_{\star}$) galaxies.  The DLA towards J1431+3952 appears to be an exception to this trend.

We finish this discussion with a review of several caveats to the mass determinations presented here, such as the adopted mass-light ratios which rely on the assumption of an early type galaxy.  Early types have effectively no dust, so using the K-band magnitude can work well as an overall magnitude estimator (K. Thanjavur 2015, private communication).  Younger galaxies have more dust and are more prone to extinction effects which can lead to an underestimate of magnitude and thus an underestimate of the mass.  The derivations were also done for a specific formation redshift, though \citet{long} mention they find similar results for different formation redshifts out to $z_f$=6. There is also significant uncertainty in the mass calculation, most of which comes from uncertainty in the stellar modelling and the availability of only K-band photometry.  The noise is also a strong contributor for calculating the flux from less luminous candidate galaxies.  Ultimately, photometric observations in additional bands are necessary to ascertain the true mass of the galaxies through spectral energy distribution calculations and spectrometry is needed for accurate redshift assesment of the candidate host galaxies.

\section{CONCLUSION}\label{conc}

The first application of ADI to direct imaging of DLA galaxies has resulted in three candidates (within 5 arcsec, or 30 kpc at the redshift of the absorber) for the $z_{abs}=0.602$ DLA seen in the quasar J1431+3952. Determination of the sensitivity curve for our observations indicates that we could have detected a galaxy whose stellar mass was as low as $10^{9.4}M_{\odot}$ at a separation of 3.4 kpc, or  $\sim 10^{9.0}M_{\odot}$ beyond $\sim$ 6 kpc. Based on the K-band photometry of our NIRI observations, we determine stellar masses of log (M$_{\star}$/M$_{\odot}$) $= 9.9, 9.7 $ and $ 11.1$ for the three candidates, which are located at impact parameters of 15, 17 and 30 kpc respectively.   The two galaxies at the lowest impact parameters are new detections in our NIRI data.  Based on a photometric redshift of $z=0.08$ (Ellison et al. 2012), the unresolved nature of the object, and inconsistency with the N(HI) -- impact parameter relation (e.g. Krogager et al. 2012; Christensen et al. 2014), we conclude that the DLA is not associated with the highest mass, largest separation object of the three candidates.  The remaining two galaxies are consistent with these scaling relations and therefore remain plausible candidates for the DLA host.  Follow-up spectroscopy is required to confirm the redshifts of the remaining two candidates and observations in just one additional band would allow for additional mass-luminosity constraints \citep{bell01}.    Our results indicate that despite its low spin temperature, the host galaxy of this DLA is unlikely to be of high stellar mass (or luminosity).  


\acknowledgments
\section*{Acknowledgments}
Based on observations obtained at the Gemini Observatory, which is operated by the Association of Universities for Research in Astronomy, Inc., under a cooperative agreement with the NSF on behalf of the Gemini partnership: the National Science Foundation (United States), the National Research Council (Canada), CONICYT (Chile), Ministerio de Ciencia, Tecnolog\'{i}a e Innovaci\'{o}n Productiva (Argentina), and Minist\'{e}rio da Ci\^{e}ncia, Tecnologia e Inova\c{c}\~{a}o (Brazil). We acknowledge and respect the native peoples on whose traditional territories the Gemini North Telescope stands and whose historical relationships with the land and Mauna a W\={a}kea (Mauna Kea) continue to this day.  Additional thanks to the comments from Luc Simard, Lise Christensen, Nissim Kanekar, Jon Willis, and Karun Thanjavur. 



\bibliographystyle{plainnat}


\end{document}